\documentclass[pdftex,12pt]{JHEP3}
\usepackage{amsmath,amsfonts,amsbsy,latexsym,amssymb,amscd,amstext}
\usepackage[pdftex]{epsfig}

\pdfoutput=1

\def\be{\begin{equation}}
\def\ee{\end{equation}}
\def\bea{\begin{eqnarray}}
\def\eea{\end{eqnarray}}
\def\pd{\partial}
\def\a{\alpha}
\def\b{\beta}
\def\g{\gamma}
\def\d{\delta}
\def\m{\mu}
\def\n{\nu}

\def\l{\lambda}

\def\s{\sigma}
\def\e{\epsilon}

\def\bi{\begin{itemize}}
\def\ei{\end{itemize}}

\date{May 25th, 2008} \preprint{IFT-UAM/CSIC-10-01\\FTUAM-10-01}
\title{ Weyl transverse gravity (WTDiff) and the cosmological constant.} \author{Enrique \'Alvarez and Roberto Vidal \\  Instituto de F\'{\i}sica Te\'orica
UAM/CSIC and Departamento de F\'{\i}sica Te\'orica \\ Universidad
Aut\'onoma de Madrid, E-28049--Madrid, Spain \\ E-mail: \email{enrique.alvarez@uam.es},\email{jroberto.vidal@uam.es}}
\abstract{
Scale invariant (transverse) gravitational theories are introduced. They are invariant under pure metric  rescalings (i.e. the matter fields are inert under those). This symmetry forbids the presence of  a cosmological constant. Those theories are not invariant under the full set of diffeomorphisms, but only with respect to those  locally characterized by the fact that their generator is transverse $\pd_\a \xi^\a=0$.
}
\begin{document}

\newpage

{\vskip 1cm}
\section{Transverse gravity and scale invariance}
The hope that  scale invariance could shed some light on the fact that  the observed value of the cosmological constant scale is much lower than expected from the wilsonian viewpoint is certainly an old and cherished one. Let us mention just a couple of  recent works \cite{Shaposhnikov:2008xb} \cite{Wetterich:2009az} where some entries into the bibliography can be found.
\par
The aim of the present work is to present a new twist of this idea in the framework of transverse gravity, where the full diffeomorphism invariance (Diff) is broken to those (TDiff) that preserve the Lebesgue measure. Transverse gravity has been studied in previous papers \cite{Alvarez:2005iy}-\cite{Alvarez:2009ga} where  references to the earlier literature are included.
\par
Those transverse gravitational models that enjoy scale invariance (that is, rigid Weyl invariance in the sense of \cite{Iorio}), dubbed WTDiff in  \cite{Alvarez:2006uu}) are (naively, as we shall see in a moment) characterized by tracefree field equations. This means that the
actions must be scale invariant, at least on shell, that is
\[
g^{\a\b}{\d\over \d g^{\a\b}}S=0
\]

The big difference with Einstein's diffeomorphism invariant gravity is that now we can sprinkle powers of $g$ here and there. Under a global (i.e. constant) Weyl rescaling
\[
g_{\a\b}\rightarrow \Omega^2 g_{\a\b}
\]
\[
g\rightarrow \Omega^{2n} g
\]
At the linear level with $\Omega\sim 1+\omega $, $\d g_{\a\b}=2\omega g_{\a\b}$.

Christoffels are invariant, and so is Riemann, so that
\[
R\rightarrow \Omega^{-2} R
\]
This means that there is a purely gravitational (without scalar fields) scale invariant \footnote{
This action can be made Weyl gauge invariant, along the lines of \cite{Iorio}
by means of a gauge field $W_\m$ that transforms as
\[
\d W_\m=\Omega^{-1}\pd_\m\Omega
\]
and adding a term proportional to
\[
\int d^n x\, |g|^{1\over n}\left(\nabla_\a W^\a + {n-2\over 2}W_\m W^\m\right)
\]
}
action, i.e.
\[
S_W\equiv -{1\over 2 \kappa^2} \int d^n x\,|g|^{1\over n} R
\]

The scaling behavior of matter is determined by the kinetic term (including the power of $|g|$ in front).

 For example, in Einstein's gravity, a scalar field with kinetic part
\[
\sqrt{|g|}{1\over 2} g^{\m\n}\pd_\m\Phi\pd_\n\Phi
\]
implies that
\[
\Phi\rightarrow \Omega^{1- n/2}\Phi
\]
which coincides with the naive dimension of the field.
\par
For Dirac fermions instead
\[
\sqrt{|g|}i\bar{\psi} e_a^\m \g^a \pd_\m\psi
\]
yields the naive dimension again
\[
\psi\rightarrow \Omega^{1-n\over 2}\psi
\]

\par
Changing the power of $|g|$, for example, as in
\[
|g|^a {1\over 2} g^{\m\n}\pd_\m\Phi\pd_\n\Phi
\]
implies that
\[
\Phi\rightarrow \Omega^{1- na}\Phi
\]
It is plain that when $a= 1/n$ then the theory enjoys rigid Weyl invariance with inert matter fields.
\par
 This means that with the measure 
 \[
 |g|^{1\over n} d^n x
 \] 
 {\em rigid Weyl invariance implies that no potential is allowed, not even a mass term.}
 \par
  Interactions are, however, allowed, but must
 either be dressed with some gravitational scalar of weight $-2$, for example,
 \[
 {c_p\over M^{p(n-2)+4-2n\over 2}} R\Phi^p
 \]
 (where $c_p$ are dimensionless constants, and $M$ is a mass scale).
 When perturbing around a nontrivial constant curvature background (such as de Sitter space), this gives rise to masses 
 \[
 m^2\sim c_2 \bar{R}
 \]
 which are naturally tiny if the radius of curvature is very large.
\par

Interacions are also allowed when they are  totally decoupled from gravitation \cite{Alvarez:2007cp}, as in
\[
d^n x\, V(\Phi_i)
\]
\par

Similarly for Dirac fermions,
\[
|g|^a i\bar{\psi} e_\a^\m \g^\a \pd_\m\psi
\]
yields
\[
\psi\rightarrow \Omega^{1-2 a n\over 2}\psi
\]
The new condition  for invariance with inert Dirac fermions is
\[
a={1\over 2n}
\]
It is remarkable that this measure does not coincide with the bosonic one.

\section{Low energy effective lagrangians}

It is expected that lowest dimension operators compatible with the assumed  symmetry (WTDiff) are bound to dominate the physics at low energies.
Let us classify transverse scalars according to their dimension, writing also the corresponding scale invariant combination.

\bi
\item {\bf Dimension zero}.

Transverse dimension zero operators are
\[
L_0\equiv F(|g|)
\]
so that the WTDiff cosmological constant is decoupled from gravity
\[
\d S_0\equiv \lambda \d \int d^n x L_0=0
\]
where $\lambda$ is be a dimension $n$ constant.
\item {\bf Dimension two}

Generic transverse dimension 2 operators are
\bea
&&L_2^{(1)}=F(|g|) g^{\a\b}\pd_\a g \pd_\b g\nonumber\\
&&L_2^{(2)}=F(|g|) R
\eea
The WTDiff operator corresponding to the first one is
\[
S_2^{(1)}\equiv -{1\over 2 \kappa_{1}^2}|g|^{1-2n\over n} g^{\a\b}\pd_\a |g| \pd_\b |g|
\]
where $\kappa_1^2$ is a new {\em gravitational constant} of dimension $2-n$ {\em a priori} unrelated to Newton's constant
\bea
&&\d S_2^{(1)}=-{1\over 2 \kappa_{1}^2}\int d^n x\left( -{1-2n\over n}|g|^{1-2n\over n} g^{\m\n}\pd_\m |g| \pd_\n |g|  g _{\a\b}\d g^{\a\b}+  |g|^{1-2n\over n}  \pd_\a |g | \pd_\b |g| \d g^{\a\b}-\right.\nonumber\\
&&\left.-2 |g|^{1-2n\over n}g^{\m\n}\pd_\m |g|\pd_\n\left(|g| g _{\a\b}\d g^{\a\b}\right)\right)=\nonumber\\
&&-{1\over 2 \kappa_{1}^2}\int d^n x\left(- {1-2n\over n}|g|^{1-2n\over n} g^{\m\n}\pd_\m |g| \pd_\n |g|  g _{\a\b}\d g^{\a\b}+|g|^{1-2n\over n} \pd_\a |g|  \pd_\b |g|  \d g^{\a\b}+\right.\nonumber\\
&&\left.+2 |g|  \pd_\n\left( |g|^{1-2n\over n}g^{\m\n}\pd_\m |g|\right) g _{\a\b}\d g^{\a\b}\right)
\eea

The gravitational equations of motion are now:
\[
 {\d S_2^{(1)}\over \d g^{\a\b}}=|g|^{1-2n\over n}\pd_\a |g|\pd_\b |g|-\left({1-2n\over n}|g|^{1-2n\over n} g^{\m\n}\pd_\m |g| \pd_\n |g|  -2|g|\pd_\n\left(|g|^{1-2n\over n} g^{\m\n} \pd_\m |g|\right)      \right)g_{\a\b}
\]
where the gravitational constant has been deleted because it is not important in the absence of matter.
These equations are  traceless {\em up to a total derivative}
\[
g^{\a\b} {\d S_2^{(1)}\over \d g^{\a\b}}=+2 n \pd_\n\left(  |g|^{1-n\over n}g^{\m\n}\pd_\m |g|\right)
\]
This means that the Noether current associated to WTDiff is
\[
W^\m\equiv   |g|^{1-n\over n}g^{\m\n}\pd_\n |g|
\]

\item {\bf Dimension two (continued)}
The second transverse dimension 2 operator is just a generalization of the usual Einstein-Hilbert lagrangian
\[
L_2^{(2)}=F(|g|) R
\]
In order to compute the variation of the corresponding WTDiff operator
\[
\d S_2^{(2)}=\d \left(-{1\over 2\kappa^2}\int d^n x |g|^{1/n} R\right)
\]
\sloppy
The variation of the curvature scalar is needed
\nopagebreak[4]
\[
\d R=\d g^{\n\s} R_{\n\s}+\left(g_{\a\b}\Delta -\nabla_{(\a}\nabla_{\b)}\right) \d g^{\a\b}
\]
\nopagebreak[4]
It follows

\bea
&&\d S_2^{(2)}=\int d^n x |g|^{1/n}\,\d g^{\a\b}\left( {1\over 2\kappa^2 n}g_{\a\b}R-{1\over 2\kappa^2}R_{\a\b}\right)\nonumber\\
&&-\int d^n x |g|^{1/n}\, {1\over 2\kappa^2}\left(g_{\a\b}\Delta -\nabla_\a\nabla_\b\right) \d g^{\a\b}
\eea

When 
\[
\d g^{\a\b}=-\Omega^2 g^{\a\b}
\]
the action remains invariant, just because $\nabla_\a g_{\m\n}=0$. We must be careful with the integration by parts.
A good place to start is the formula valid for any contravariant vector \cite{Eisenhart}
\[
\nabla_\m V^\m={1\over \sqrt{|g|}}\pd_\m\left(\sqrt{|g|} V^\m\right)
\]
Let us integrate by parts the slightly more general integral
\bea
&&\int d^n x f(g)\,\nabla_\m \left(\nabla^\m g_{\a\b} \d g^{\a\b}-\nabla_\b\d g^{\m\b}\right)\equiv I_1-I_2\nonumber\\
&&I_1\equiv\int \pd_\m\left({f\over \sqrt{|g|}}\right)\sqrt{|g|}\nabla_\b\d g^{\m\b}=\int \pd_\m\left({f\over \sqrt{|g|}}\right) \sqrt{|g|}\left(\pd_\b \d g^{\m\b}+\Gamma^\m_{\b\s}\d g^{\s \b}+\Gamma^\b_{\b\s}\d g^{\s\m}\right)=\nonumber\\
&&\quad=\int \pd_\m\left({f\over \sqrt{|g|}}\right) \sqrt{|g|}\left(\Gamma^\m_{\b\s}\d g^{\s \b}+\Gamma^\b_{\b\s}\d g^{\s\m}\right)
-\pd_\b\left(\pd_\m\left({f\over \sqrt{|g|}}\right)\sqrt{|g|}\right) \d g^{\m\b}\nonumber\\
 &&I_2\equiv \int \pd_\m\left({f\over \sqrt{|g|}}\right)\sqrt{|g|}\nabla^\m g_{\a\b} \d g^{\a\b}= \int \pd_\m\left({f\over \sqrt{|g|}}\right)\sqrt{|g|}g^{\m\l}\pd_\l\left( g_{\a\b} \d g^{\a\b}\right)=\nonumber\\
&&\quad=-\int\pd_\l \left( \pd_\m\left({f\over \sqrt{|g|}}\right) \sqrt{|g|}
g^{\m\l}\right)  g_{\a\b} \d g^{\a\b}
\eea
\nopagebreak
In conclusion\footnote{
It is worth checking that this still gives zero for a metric rescaling. This means that both integrals must vanish separately  when
\[
\d g^{\a\b}= -\e g^{\a\b}
\]
This is obvious for $I_2$, which in this case reduces to the integral of a total derivative. With respect to the first integral, we shall employ  the well-known formulas \cite{Eisenhart}
\bea
&&\Gamma^\b_{\b\s}={1\over\sqrt{|g|}}\pd_\s \sqrt{|g|}\nonumber\\
&&g^{\a\b}\Gamma^\m_{\a\b}=-{1\over\sqrt{|g|}}\pd_\l\left(\sqrt{|g|} g^{\m\l}\right)
\eea
relating the Christoffels and the determinant.
\bea
&&I_1=\int \Phi_\m\sqrt{|g|}\left(-{1\over\sqrt{|g|}}\pd_\l\left(\sqrt{|g|} g^{\m\l}\right)+{1\over\sqrt{|g|}}g^{\s\m} \pd_\s \sqrt{|g|}\right)-g^{\m\b}\pd_\b\left(\sqrt{|g|}\phi_\m\right)=\nonumber\\
&&=\int \sqrt{|g|}\Phi_\m\pd_\b g^{\m\b}-\Phi_\m \sqrt{|g|}\pd_\l g^{\m\l}-\Phi_\m g^{\m\l}\pd_\l \sqrt{|g|}+\Phi_\m g^{\s\m}\pd_\s \sqrt{|g|}=0
\eea
That is, the integrand itself vanishes.
Under an arbitrary variation
\bea
&&{\d I_1\over \d g^{\a\b}}=\Phi_\m \sqrt{|g|}\Gamma^\m_{\a\b}+\Gamma^\l_{\l\a}\phi_\b \sqrt{|g|}-\pd_\b\left(\sqrt{|g|}\Phi_\a\right)=\nonumber\\
&&-\sqrt{|g|}\pd_\b\Phi_\a+\Phi_\m \sqrt{|g|}\Gamma^\m_{\a\b}\equiv-\sqrt{|g|}\nabla_\b\Phi_\a\nonumber\\
&&{\d I_2\over \d g^{\a\b}}=-\pd_\l\left(\Phi_\m\sqrt{|g|} g^{\m\l}\right)g_{\a\b}\equiv - \sqrt{|g|}\nabla_\l\left(\Phi_\m g^{\m\l}\right)g_{\a\b}
\eea
where the covariant derivatives are defined as if $\Phi_\m$ were a tensor; which it is not, so that those constructions do not enjoy all properties of covariant derivatives of tensors. Still, it is sometimes a useful abbreviation. 
}, calling $\Phi_\mu=\pd_\m\left({f\over \sqrt{|g|}}\right)$\pagebreak
\bea
&&\d S_2^{(2)}=\int d^n x |g|^{1/n}\,\d g^{\a\b}\left( {1\over2\kappa^2  n}g_{\a\b} R-{1\over 2\kappa^2}R_{\a\b}\right)+\nonumber\\
&&+\int d^n x  {1\over 2\kappa^2}\sqrt{|g|}\left(\nabla_{\b}\Phi_{\a}-\nabla_\l\left(\Phi_\m g^{\m\l}\right)g_{\a\b}\right)\delta g^{\a\b}=\nonumber\\
&&=\int |g|^{1/n} {1\over 2\kappa^2}\Bigg[\left( {1\over n}g_{\a\b} R-R_{\a\b}\right)+{2-n\over 2n}|g|^{-1}\Bigg({2-3n\over 2n}g^{-1}g_\a g_\b\qquad\qquad\nonumber\\
&&\qquad\qquad\quad-\nabla_\b g_\a-\left({1-n\over n}g^{-1}g_\m g_\n g^{\m\n}+\partial_\m(g_\n g^{\m\n})\right)g_{\a\b}\Bigg)\Bigg]\d g^{\a\b}d^n x
\eea

It is remarkable that Einstein's 1919 equations
\[
R_{\m\n}-{1\over n} R g_{\m\n}= \kappa^2 \left(T_{\m\n}-{1\over n}T g_{\m\n}\right)
\]
which are truly traceless
 \cite{Einstein}, cf. also  \cite{Alvarez:2005iy} do not seem to be obtainable from a variational principle of the sort we are studying which always yield equation of motion which are traceless only up to a total derivative.

\ei

\section{Conclusions}
We have studied in the body of the paper a gravitational symmetry that forbids the presence of a cosmological constant. We believe that this is some progress insofar as we were not aware of any such symmetry previously known.
\par
It would be interesting to present our results in the Einstein frame. In the case of the second dimension 2 operator, which is the only one resembling the Einstein-Hilbert lagrangian, this would stem from the redefinition of a new spacetime metric such that
\[
\sqrt{|g_e|}R[g_e]=|g|^{1\over n}R
\]
It is quite simple to realize that
\[
g^e_{\m\n}\equiv g^{-{1\over n}}g_{\m\n}
\]
such that $g_e\equiv 1$. The restricted variational principle would then give true traceless equations of motion of the Einstein's 1919 sort \cite{Einstein}, except that in Einstein's mind the metric was not restricted by any unimodularity condition.
\par
We can understand our results from a different viewpoint. It is well known that transverse theories are equivalent, in a given reference system, to scalar-tensor theories
\cite{Buchmuller:1988wx}\cite{Alvarez:2006uu}. A way of implementing this mapping is as follows: our second dimension 2 lagrangian is equivalent to 
\[
L=-{1\over 2\kappa^2}\sqrt{|g|}\phi R+\sqrt{|g|}\chi\left(\phi-|g|^{2-n\over 2n}\right)
\]
where $\phi$ and $\chi$ are two auxiliar scalar densities. It is now possible to find an unconstrained Einstein metric such that
\[
\sqrt{|g_E|}R[g_E]=\sqrt{|g|}\phi R
\]

The answer is clearly
\[
g^E_{\m\n}=\phi^{2\over n-2} g_{\m\n}
\]
(so that $g_E=g \phi^{2n\over n-2}$) and the full scalar-tensor lagrangian reads

\bea
&&L=-\frac1{2\kappa^2}\phi\sqrt{|g|}R+\sqrt{|g|}\chi(\phi-|g|^{2-n\over 2n})=-\frac1{2\kappa^2}\sqrt{|g_E|}R_E+\sqrt{|g_E|}\phi^{-\frac2{n-2}}\chi(1-|g_E|^{2-n\over 2n})+\nonumber\\
&&+\frac{n-1}{2\kappa^2(n-2)}\left(2\partial_\mu\left(\sqrt{g_E}g_E^{\mu\nu}\frac{\pd_\n\phi}{\phi}\right)-\sqrt{g_E}g_E^{\m\n}\frac{\pd_\m\phi\pd_\n\phi}{\phi^2}\right)\eea
id est, it is of the unimodular type. It has however been stressed in the literature \cite{Alvarez:2005iy} that this is subtly not equivalent to choosing the unimodular gauge in general relativity, which is always allowed (and used many times by Einstein himself).
The coupling to matter is independent of the scalar density $\phi$. For example, for a scalar field  $\Phi$ (not to be confused with the scalar density $\phi$ of gravitational origin),
\[
L_I=|g_E|^{1\over n} g_E^{\m\n} \pd_\m\Phi \pd_\n \Phi
\]
\par
Under conformal transformations in the old frame
\[
\phi\rightarrow \Omega^{2-n}\phi 
\]
and for consistency,
\[
\chi\rightarrow \Omega^{-2}\chi
\]
whereas the unimodular Einstein metric is inert. What looks like a purely gravitational symmetry in one frame, looks like a {\em matter} symmetry in another. Potential energy coupled to gravitation is again forbidden, because they appear in the new frame as
\[
\phi^{-{2\over n-2}}V(\Phi)
\]
\par
It is also interesting to follow the first dimension 2 term under this change of frame. It is easy to check that if the equations of motion are used, it reduces to
\[
L_2^{(1)}={4 n^2\over (n-2)^2}\phi^{-2} g_E^{\m\n}\pd_\m \phi \pd_\n\phi
\]
(if the equations of motion are not used, there are other terms proportional to $\pd_\m |g_E|$).
\par
 Nevertheless, transverse theories are most likely severely constrained by experiment \cite{Alvarez:2009ga} and besides scale invariance has to be broken, at least by the Weyl anomaly \cite{Duff:1993wm}\cite{Boulanger:2007ab} (which has yet to be computed for transverse theories).
 \par
 Actually WTDiff makes an overkill, in the sense that it not only forbids a cosmological constant, but also any potential energy whatsoever which is coupled to gravitation. There is experimental evidence\footnote{
 Although experiments tend to bound {\em differences} between properties of different objects, so that if those differences are universal there are not so constrained.}
 that potential energy does couple to gravitation \cite{Carlip:1997gf}, which is again an indication that scale symmetry must be badly broken in nature.
 \par
 The proper setting of the problem is most likely a cosmological one, in which the universe goes through different epochs characterized by different amounts of symmetry in the gravitational sector. Work on concrete models of this sort is in progress and we hope to report on that in the future.

\section*{Acknowledgments}
One of us (EA) is grateful from stimulating discussions with Jaume Garriga, Diego Blas, Alex Kehagias, Oriol Pujolas  and Enric Verdaguer.
This work has been partially supported by the
European Commission (HPRN-CT-200-00148) as well as by FPA2009-09017 (DGI del MCyT, Spain) and 
 S2009ESP-1473 (CA Madrid). R.V. is supported by a MEC grant, AP2006-01876.

               

\begin{thebibliography}{99}
\bibitem{Alvarez:2005iy}
  E.~Alvarez,
  ``Can one tell Einstein's unimodular theory from Einstein's general
  relativity?,''
  JHEP {\bf 0503}, 002 (2005)
  [arXiv:hep-th/0501146].

\bibitem{Alvarez:2006uu}
  E.~Alvarez, D.~Blas, J.~Garriga and E.~Verdaguer,
  ``Transverse Fierz-Pauli symmetry,''
  Nucl.\ Phys.\  B {\bf 756} (2006) 148
  [arXiv:hep-th/0606019].
\bibitem{Alvarez:2007cp}
  E.~Alvarez and A.~F.~Faedo,
  ``A comment on the matter-graviton coupling,''
  Phys.\ Rev.\  D {\bf 76} (2007) 124016
  [arXiv:0707.4221 [hep-th]].\\
 E.~Alvarez and A.~F.~Faedo,
  ``Unimodular cosmology and the weight of energy,''
  Phys.\ Rev.\  D {\bf 76} (2007) 064013
  [arXiv:hep-th/0702184].
  
\bibitem{Alvarez:2009ga}
  E.~Alvarez, A.~F.~Faedo and J.~J.~Lopez-Villarejo,
  ``Observational constraints on transverse gravity,''
  JCAP {\bf 0907} (2009) 002
  [arXiv:0904.3298 [hep-th]].


\bibitem{Boulanger:2007ab}
  N.~Boulanger,
  ``Algebraic Classification of Weyl Anomalies in Arbitrary Dimensions,''
  Phys.\ Rev.\ Lett.\  {\bf 98} (2007) 261302
  [arXiv:0706.0340 [hep-th]].

\bibitem{Buchmuller:1988wx}
  W.~Buchmuller and N.~Dragon,
  Phys.\ Lett.\  B {\bf 207} (1988) 292.

\bibitem{Carlip:1997gf}
  S.~Carlip,
  ``Kinetic Energy and the Equivalence Principle,''
  Am.\ J.\ Phys.\  {\bf 65} (1998) 409
  [arXiv:gr-qc/9909014].


\bibitem{Deser:1993yx}
  S.~Deser and A.~Schwimmer,
  ``Geometric classification of conformal anomalies in arbitrary dimensions,''
  Phys.\ Lett.\  B {\bf 309} (1993) 279
  [arXiv:hep-th/9302047].
  
\bibitem{Duff:1993wm}
  M.~J.~Duff,
  Class.\ Quant.\ Grav.\  {\bf 11} (1994) 1387
  [arXiv:hep-th/9308075].



\bibitem{Einstein}
A. Einstein, Siz. Preuss. Acad. Scis. 1919, english translation in The principle of relativity,
A. Einstein et al. eds., Dover.

\bibitem{Eisenhart}
L.P. Eisenhart, 
"Riemannian geometry"
(Princeton University Press)


\bibitem{Fefferman}
C. Fefferman and C.R. Graham,
"Conformal invariants"
Asterisque hors s\'erie,1985,p.95-116.

\bibitem{Iorio}
  A.~Iorio, L.~O'Raifeartaigh, I.~Sachs and C.~Wiesendanger,
  ``Weyl gauging and conformal invariance,''
  Nucl.\ Phys.\  B {\bf 495} (1997) 433
  [arXiv:hep-th/9607110].
\bibitem{Shaposhnikov:2008xb}
  M.~Shaposhnikov and D.~Zenhausern,
  ``Scale invariance, unimodular gravity and dark energy,''
  Phys.\ Lett.\  B {\bf 671} (2009) 187
  [arXiv:0809.3395 [hep-th]].
  
\bibitem{Wetterich:2009az}
  C.~Wetterich,
  ``The cosmological constant and higher dimensional dilatation symmetry,''
  arXiv:0911.1063 [hep-th].


  
  
  
  
  
  
  
  
  
  
  
  
  
  
  
  
  
  
  
  
  
  
  

\end{thebibliography}
\end{document}